\documentclass[aps,prl,twocolumn,groupedaddress]{revtex4-1}

\usepackage[pdftex]{graphicx}

\begin{document}

\title{Characterizing the Rheology of Fluidized Granular Matter}

\author{Kenneth W. Desmond$^1$, Umberto Villa$^2$, Mike Newey$^3$, Wolfgang Losert$^3$}
\email[]{wlosert@umd.edu}
\affiliation{$^1$ Department of Physics, Emory University, Atlanta, GA\\ 
		$^2$ Department of Mathematics \& Computer Science, Emory University, Atlanta, GA\\ 
		$^3$Department of Physics, IPST, and IREAP, University of Maryland, College Park, MD}

\date{\today}

\begin{abstract}
In this study we characterize the rheology of fluidized granular matter subject to secondary forcing.  Our approach consists of first fluidizing granular matter in a drum half filled with grains via simple rotation, and then superimposing oscillatory shear perpendicular to the downhill flow direction. The response of the system is mostly linear, with a phase lag between the grain motion and the oscillatory forcing. The rheology of the system can be well characterize by the GDR-Midi model if the system is forced with slow oscillations.  The model breaks down when the forcing timescale becomes comparable to characteristic time for energy dissipation in the flow.

\end{abstract}


\maketitle

Describing and accurately predicting the flow of granular matter is important for many industrial applications, as well as geological processes~\cite{Lueptow2002,JaegerRevModernPhys1996,ScottNature1996,Duran2000}.   For example, accurate models for the flow of dense granular matter are important for applications such as soil compaction~\cite{Bolton1979}, rock avalanches~\cite{Losert2003, Losert2005}, or manufacturing of pills from powders~\cite{deGennesRevModernPhys1999}.  
However, unlike fluids where flow is well described by the Navier-Stokes equations, no such set of equations has been shown to predict the large variety of observed granular flows.  Nevertheless granular flows tend to be very reproducible and describeable by simple empirical continuum equations.  

Recently significant progress has been made to classify a wide range of granular flow geometries within a unifying empirical framework.  da Cruz \textit{et al.} and the GDR-Midi group have developed a model employing
a single dimensionless parameter, the inertia number, that characterizes the flow behavior based on the
importance of inertia relative to the confining pressures~\cite{Cruz2005, GDRMidi, Jop2006, Delannay2007}. Key aspects of the GDR-MiDi model are the presence of a yield stress and the approach of an effective viscosity from a non-zero constant towards a larger asymptotic value in the limit of large strain rate.  The GDR-Midi model has been successfully used to develop constitutive equations that accurately predict shear flow profiles for various geometries such as chute flow and rotating drum flow. 

Test of this model and other models of granular flow~\cite{Forterre2008} have focused on 2D flow profiles.  In such flows, from a microscopic view of forces transmitted through particle contacts, the fabric tensor and force chains ~\cite{Behringer2004} maintain direction or rotate within a plane.  Since the GDR-Midi model was developed with such flow profiles it is likely that this effect is contained in the expression for the effective viscosity.  However, many relevant boundary conditions in granular flows involve additional forces perpendicular to the 2D flow profile and 3D flow fields. Common examples include rock avalanches flowing down real surface topographies, or grains exiting the orifice of a hopper.  From the microscopic view the fabric tensor and force chains are now forced in 3D rather than a plane.   

\begin{figure}
\includegraphics[width=3.25in]{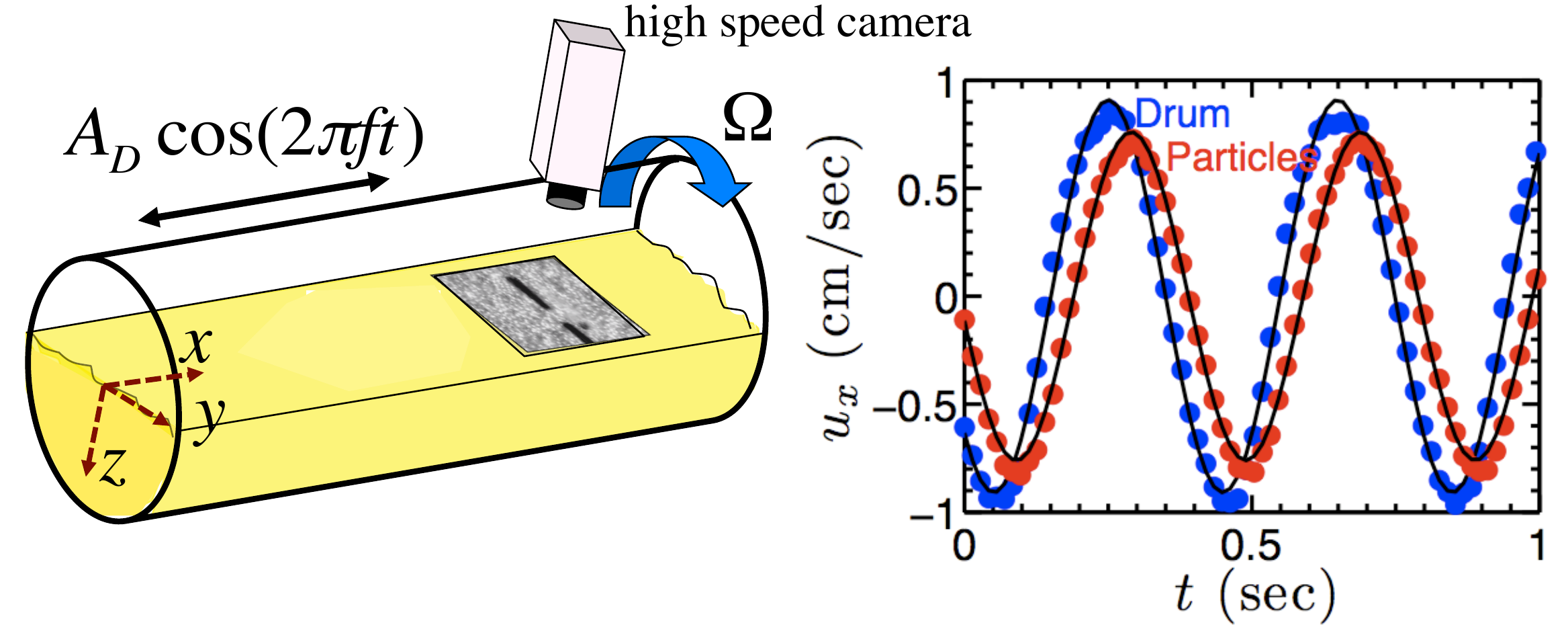}
\caption{(Color online.) Schematic of the experimental apparatus. The drum is half filled with glass beads of diameter $d$ and density $\rho$, and the grains are fluidized by rotating the drum with angular frequency $\Omega$. A secondary shear is applied by oscillating the drum with frequency $f$, and the oscillating motion of the grains at the surface is recorded with a high speed camera. The plot to the right shows the motion of the drum (dark, blue) and 2mm grains (light, red) measured for $f= 2.5 $ Hz; plus sinusoidal fits (solid lines).}
\label{fig:Schematic}
\end{figure}

In this paper, we present a new table top geometry to characterize the rheology of granular matter that is already exhibiting a shear flow, with the aim to elucidate how flowing granular matter responds to additional forcing.  Our approach consists of first fluidizing granular matter by rotation in a horizontal cylinder rotated about the cylindrical axis, as shown in the schematic in Fig.~\ref{fig:Schematic}. In addition to rotation, shear is applied orthogonal to the flow direction via horizontal oscillation of the entire cylinder. To test the GDR-Midi model we compare our results to the prediction of the model. We find that the model is unable to predict the rheology at large, fast forcing, but is able to reproduce the correct scaling relationship and predict the rhoelogy at low forcing.

In our experiment, we rotate a drum of diameter 10 cm and length 67 cm half full of monodisperse beads about the long axis at frequency $\Omega$. We use glass beads (Glenn Mills, Inc.) with diameter $d$.
A high speed camera (Photron FastCam) placed above the drum records the motion of the grains at a frame rate of 500 frames/sec for 2.096 seconds. The camera is located 10 cm from the end of the cylinder to avoid boundary effects from the end caps.

To assess the rheology of the flowing layer a secondary forcing is superimposed by driving the drum sinusoidally along the long axis at driving frequency $f$ (see Fig.~\ref{fig:Schematic}). We define the $x$-axis to be parallel to the oscillating motion and the $z$-axis to be perpendicular to the granular surface.

Using images taken with the high speed camera both the motion of the grains and the drum are tracked simultaneously using an algorithm by Crocker \textit{et al.}~\cite{Crocker1996}. The black stripes seen in the middle of the image is tape placed at the surface of the drum to identify the drum's horizontal position. Figure~\ref{fig:Schematic} shows a plot of both the horizontal $x$-axis motion of grains at the surface and the horizontal motion of the drum, respectively.

\begin{figure}
\includegraphics[width=3.25in]{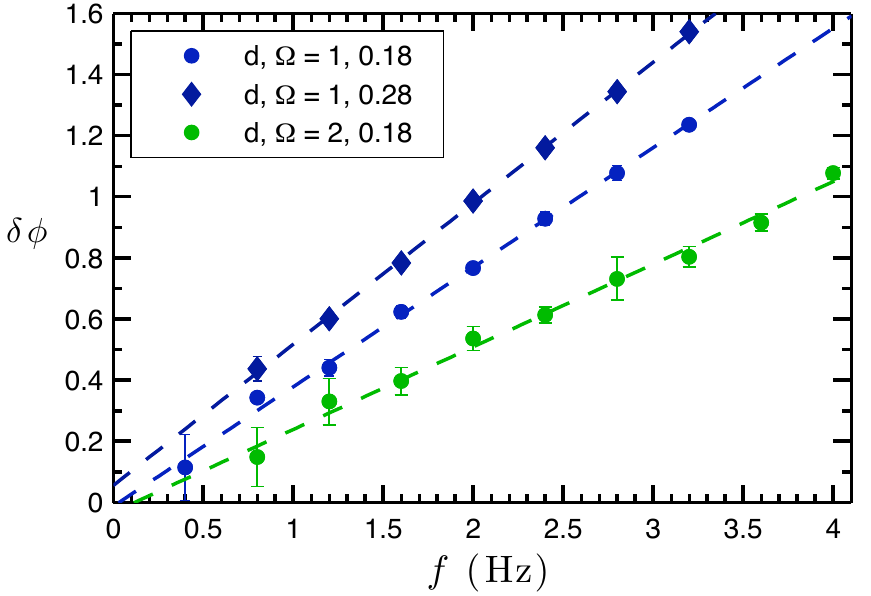}
\caption{(Color online.) (a) Phase lag $\delta{}\phi{}(f)$ vs oscillation frequency for representative particle sizes $d$ and rotation rates $\Omega$. The units for $d$ and $\Omega$ are mm and Hz, respectively.}
\label{fig:omegadependence}
\end{figure}

The velocity of the drum and the grains along the drum axis in the $x$-direction $u_x$ 
(see Fig.~\ref{fig:Schematic}) are fit to 
a sinusoidal function $u_x = -2\pi{}Af\sin(2\pi{}ft + \phi)$
to obtain a frequency $f$, amplitude $A$, and phase $\phi$ of the particle and drum motion.  The phase of particle motion $\phi_P$ and phase of drum motion $\phi_D$ differ.  The peak acceleration due to horizontal shaking is less than $0.02{\rm g}$, significantly smaller than the downhill acceleration of particles and too small to fluidize the material directly when the tumbler is not rotated. However, for our larger oscillation frequencies near 4 Hz the downhill velocity becomes slightly perturbed.

The frequency ($f$) dependent response of the fluidized granular matter to oscillatory forcing can 
be described with the phase lag $\delta{}\phi = \phi_D - \phi_P$. Figure~\ref{fig:omegadependence}(a) shows the frequency dependence for a few representative values of $d$ and $\Omega$. The data shows that the phase lag varies linearly with oscillation frequency $f$ and goes to zero as $f \rightarrow 0$ within experimental error. 

We note that an oscillating layer of Newtonian fluid would exhibit a scaling of $\delta\phi \sim \sqrt{f}$, unlike the linear scaling we observe in our data~\cite{LandauFluidMechanics}.  This discrepancy implies that in order to accurately model the 
behavior of this system a non-newtonian model is needed. 

Thus we apply the GDR-Midi model that has been successfully used to model simple tumbler flows. 
The GDR-Midi model is a continuum model treating the granular medium as a frictional material with a shear rate and pressure dependent friction coefficient.  We simulate the model as layers of solid planes slipping past each other and interacting through friction, similar to a da Vinci fluid model~\cite{Blumenfeld2010}. When there is a velocity difference between adjacent layers they exhibit stresses on each other $\sigma = \mu{}P\text{sign}(\dot{\gamma})$, where $\mu$ is a strain rate dependent friction coefficient, $P$ is the pressure, and $\dot{\gamma}$ is the strain rate. The strain rate dependence of $\mu$ is written as 
\begin{equation}
\mu = \mu_s + (\mu_2 - \mu_s)/(1 + I_0/I),
\label{eq:mu}
\end{equation} 
where $\mu_s$, $\mu_2$, and $I_0$ are material dependent constants, and the inertia number $I = |\dot{\gamma}|d/\sqrt{P/\rho}$, where $d$ is the particle diameter and $\rho$ is the particle density. This model can be recast into a Non-Newtonian fluid model by defining an effective viscosity as $\eta = \mu(I)P/|\dot{\gamma}|$ such that $\sigma = \eta{}\dot{\gamma}$. 

To reduce the complexity of our system we will treat the rotating drum apparatus as a oscillating inclined plane. In this geometry, the GDR-Midi model can be written as the following set of coupled differential equations
\begin{eqnarray}
\label{eq:GDRModel}
&& \rho{}\frac{\partial{}u_y(z)}{\partial{}t} = \frac{\partial}{\partial{}z}\left(\eta(\dot{\gamma}, P)  u_{y,z} \right)  + g \sin(\theta) \nonumber\\
&& \rho{}\frac{\partial{}u_x(z)}{\partial{}t} = \frac{\partial}{\partial{}z}\left(\eta(\dot{\gamma}, P) u_{x,z} \right) \label{eq:diffeq} \\
&& \dot{\gamma}_{ij} = u_{i,j} + u_{j,i} \ \  \text{\&} \ \  |\dot{\gamma}| =  \sqrt{\frac{1}{2}\left(  tr(\dot{\gamma}_{ij} )^2-tr(\dot{\gamma}_{ik}\dot{\gamma}_{kj}) \right)} \nonumber
\end{eqnarray}
where the notation $u_{i, j}$ is the partial derivation of the velocity in the $i$ direction with respects to the $j$ direction, $\dot{\gamma}$ is the strain tensor, and $\theta$ is the angle of inclination of the surface. The first equation governs the flow down the inclined plane and the second equation governs the flow in the oscillatory direction. In the absence of secondary forcing (typical incline plane flow) Jop \textit{et al.}~\cite{Jop2005_2} showed that $\mu(z)$ is constant with depth in the steady state. As a consequence, the inertia number is also a constant independent of $z$ where $I = [(\mu_2 - \mu_s)/(\mu - \mu_s)-1]^{-1}I_o$. Using the definition of inertia number and $P = \rho{}gz\cos(\theta)$, one can invert the previous equation to find $\dot{\gamma{}}(z) = C\sqrt{z}$, where the constant $C = I/(d\sqrt{g\cos(\theta)})$. Therefore, in the absence of secondary forcing there exist an analytical form describing the flow. For the fully coupled equations shown in Eqn.~\ref{eq:GDRModel} there is no analytical solution. Instead we must use numeric integration to solve for the flow profile, and we have checked that our numerical integration reduces to the correct solution in the absence of oscillatory flow.

Before we continue to the results of the model we seek to understand if there is a natural time scale $t_{\eta}$ present in the system so that we can compare our forcing time scale $t_{f} = 1/f$ to the relaxation time of the material. To find $t_{\eta}$ we consider the differential equation describing the downhill flow in the absence of secondary forcing. In this differential equation $\eta$ can be written explicitly as a function of $z$ using our results from before, where $\eta = \mu{}\rho{}g\cos(\theta)\sqrt{z}/C$. Making a change of variable $z' = z/h$, where $h$ is the thickness of the flowing layer, and plugging $\eta$ into the differential equation for the downhill flow we find 
\begin{equation}
\frac{\partial{}u_y(z')}{\partial{}t} = \frac{\mu{}g\cos(\theta)}{Ch^{3/2}}\frac{\partial}{\partial{}z'}\left( \sqrt{z'}u_{y,z'} \right)  + g \sin(\theta).
\label{eq:timescale}
\end{equation} 
By dimensional arguments the above equation has a characteristic time scale $t_{\eta} = Ch^{3/2}/(\mu{}g\cos(\theta))$, where $t_{\eta}$ is a characteristic time scale for the flowing layer to dissipate it's energy. For instance, if we consider the case of a flowing granular layer of thickness $h$ at an inclination angle $\theta$, $t_{\eta}$ is a characteristic time for the grains to stop flowing if we suddenly changed the inclination angle to $\theta = 0^{\circ{}}$. We also note that by integrating the strain rate, the downhill velocity at the top of the flowing layer is $u_y(z =  0) = u_{top} = 2/3Ch^{3/2}$ and therefore $t_{\eta} = 3u_{top}/(2\mu{}g\cos(\theta))$.

Using the two time scales $t_{\eta}$ and $t_{f}$ we can define a dimensionless number $t_{\eta}/t_{f}$ which characterizes the influence of the secondary forcing. When $t_{\eta}/t_{f} \ll 1$ the secondary forcing takes place over a much longer time scale than the relaxation time of the grains and therefore during each oscillation the grains can relax. However, when $t_{\eta}/t_{f} \gtrsim 1$ the secondary forcing is fast enough that during each oscillation the grains will not be able to completely relax. We may expect that as $t_{\eta}/t_{f}$ approaches unity the response of the material may change.

\begin{figure}
\includegraphics[width=3.25in]{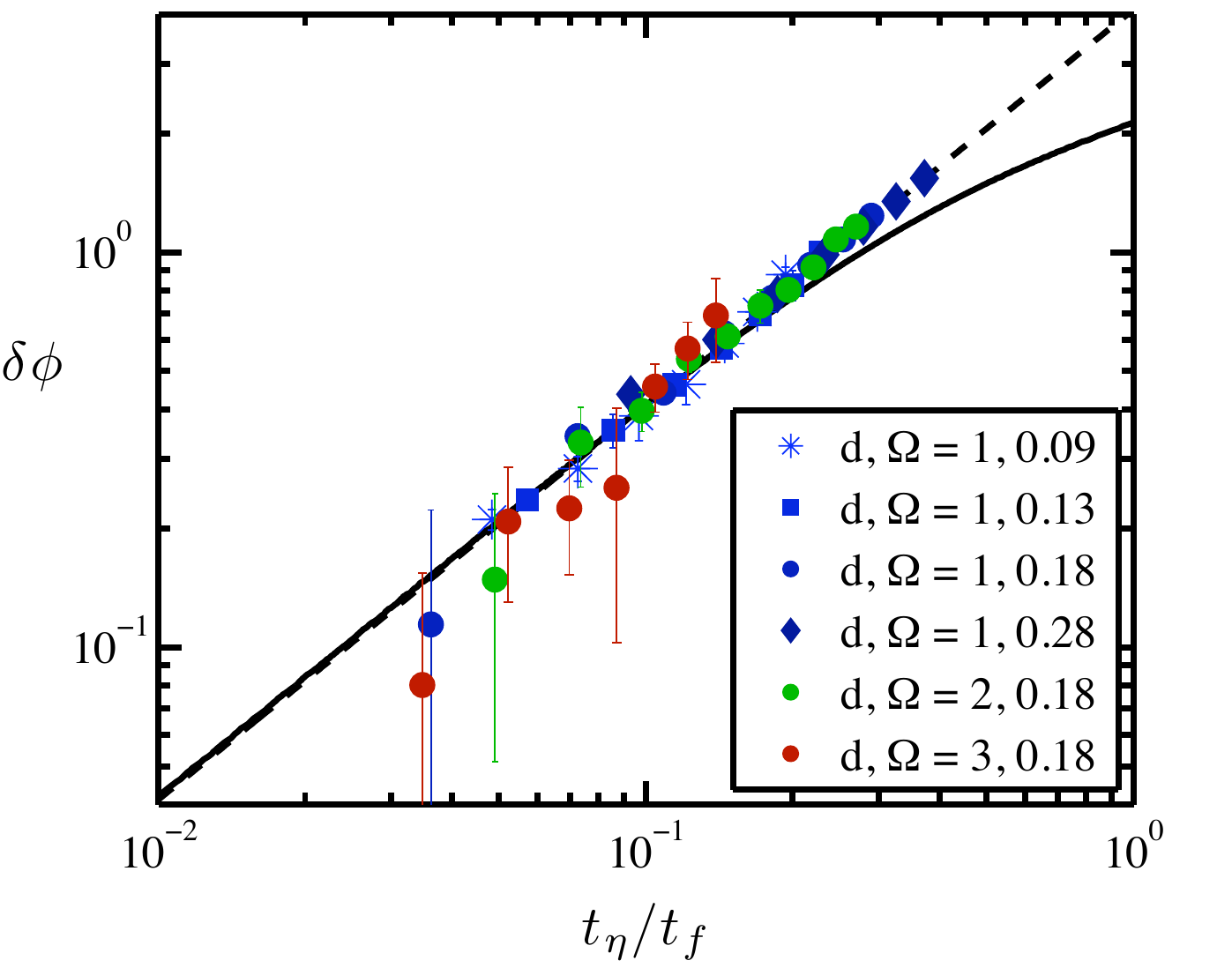}
\caption{(Color online.) Dimensionless response curves are shown for our model and the experimental data. The black solid line is data from the model and the black dash line is an extension of the linear regime found in the model and is meant as a guide to the eye. The symbols are experimental data. The units for $d$ and $\Omega$ are mm and Hz, respectively.}
\label{fig:scalingdata}
\end{figure}

Now knowing the dimensionless parameter describing the system we numerically integrate the coupled differential equations Eqn.~\ref{eq:GDRModel} to solve for over 800 different $\delta{}\phi{}(f)$ response curves for various combinations of parameters spanning the ranges: $d = 1-3$ mm, $\theta = 18^o - 28^o$, $h = 7 - 15$ mm, $\mu_s = 0.33 - 0.42$, $\mu_2 = 0.55 - 0.64$ and $I_o = 0.279$. The results of the model are shown in Fig.~\ref{fig:scalingdata} as the black solid line, where the response curves are plotted as a function of $t_{\eta}/t_{f}$. The black line shows that all 800 response curves collapse to a single curve when rescaled by the dimensionless ratio $t_{\eta}/t_{f}$, indicating that $t_{\eta}/t_{f}$ is an appropriate measure to characterize the influence of the secondary forcing. The data have been plotted on a log-log scale to show that at small $t_{\eta}/t_{f}$ the model predicts a linear response regime. 

To compare our experimental results to the model we must compute $t_{\eta}/t_{f} = 3u_{top}f/(2\mu{}g\cos(\theta))$ for our experimental data. However, for our experiments we do not know the value of $\mu$, though we do know the other values $\theta$, $g$, $u_{top}$, and $f$. Therefore we treat $\mu$ as a fitting parameter and find $\mu$ for each experiment by fitting $\delta{}\phi(f)$ to the linear response regime found from the model. The experimental results are shown as symbols in Fig.~\ref{fig:scalingdata}, where each data set was fitted to the black dash line to find a value for $\mu$. The fitted values of $\mu$ ranged from 0.45 - 0.54.  In the GDR-Midi model for steady state incline flow $\mu = \tan(\theta)$~\cite{Jop2005_2}. For our experiments, $\theta \sim 23^{\circ}$ giving an expected value of $\mu \sim 0.43$ close to the values found for our experiments.

Figure ~\ref{fig:scalingdata} shows that the linear response regime of the experiment extends beyond the linear response regime predicted by the GDR-Midi model. Since the experimental data at larger $t_{\eta}/t_{f}$ have a larger phase lag than predicted, it suggest that the effective viscosity of a real granular material upon large secondary forcing is smaller than the model predicts.  One possibile reason why the material exhibits less flow resistance under fast forcing may be that granular flows are transliently weak when their principal stress axis is changed~\cite{Toiya2004}, likely since finite strain is needed to realign the fabric tensor and force chains in a new direction.  Our experimental results indicate that the material remains weak up to a timescale comparable to the visous relaxation timescale.  

\begin{figure}
\includegraphics[width=3.25in]{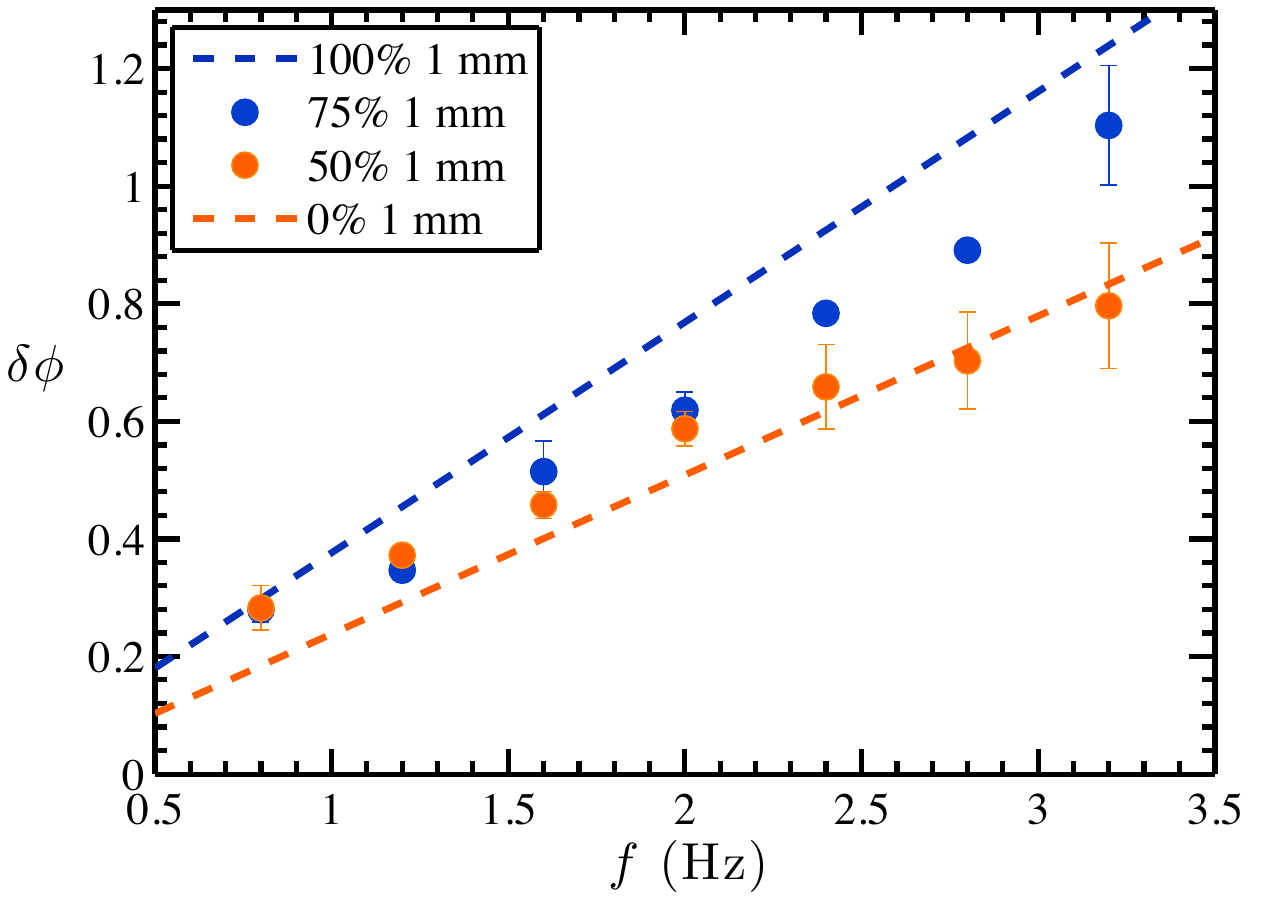}
\caption{(Color online.) $\delta{}\phi$ is shown for two different binary mixtures of 1 mm and 2 mm particles. The blue (dark gray) data points are for a mixture of 75\% 1 mm beads by volume and the orange (light gray) data points are for a mixture of 50\% 1 mm by volume. Data for a the monodisperse systems are also shown.}
\label{fig:mixture}
\end{figure}

With our new method now established, we also investigate the behavior of two binary mixtures of 1 mm and 2 mm diameter particles. Figure~\ref{fig:mixture} shows the response curve for two different mixture ratios of 1 mm to 2 mm beads at $\Omega = 0.18$ Hz. To ensure that the grains had fully segregated we allowed the drum to rotate at least 50 times before recording the flow.

The curves in Fig.~\ref{fig:mixture} show that the mixtures behave in a manner that is bound between the behavior observed for pure 1 mm  or 2 mm bead systems, and exhibit a non-linear response with increasing $f$ which is much different than the monodisperse system and the GDR-Midi model. We find that for small $f$ both systems behave like a system of pure 1 mm beads. This is surprising given that after segregation very few 1 mm particles will be in the fluidized layer, but yet these few 1 mm particles can significantly weaken a 2 mm sample. However, for higher $f$ we see that the system containing 50\% 1 mm beads converges to the pure 2 mm bead curve, while the system containing 75\% 1 mm particles has a response that is between the behavior observed for the 1 mm and 2 mm monodisperse systems. This suggests that with increasing oscilation frequency more 1 mm beads may be involved in the oscillatory flow field.  

In this paper we have introduced a new approach to quantify the response of fluidized grains to secondary forcing. Our new apparatus consist of first fluidizing granular matter by rotation in a horizontal cylinder rotated about the cylindrical axis, as shown in the schematic in Fig.~\ref{fig:Schematic}. In addition to rotation, shear is applied orthogonal to the fluidizing mechanism via horizontal oscillation of the entire cylinder to probe the rheological response of the fluidized grains. We find that fluidized monodisperse spheres at low forcing are well described by the GDR-Midi model and that the response function $\delta{}\phi{}(f)$ can be used to find the effective friction coefficient $\mu$ for steady state shear. However, at faster forcing frequencies the GDR-Midi model breaks down, predicting a larger effective viscosity than observed in the experiment. We attribute the smaller than expected viscosity observed in experiments to the slow adaptation of the fabric tensor and force chain directions to the direction of forcing.  Unlike traditional rheometers that elucidate materials properties from the amplitude and phase of the response to oscillatory forcing, we suggest that the rheology of the flowing state is best determined via small perturbations
of a primary flow field.  Our device thus provides a template for a simple rheometer for the flowing state of granular matter.

%%%%%%%%%%%%%%%%%%%%%%%%%%%%%%%%%%%%%%%%%%%%%%%
%%%%%%%%%%%%%%%%%%%%%%%%%%%%%%%%%%%%%%%%%%%%%%%

\begin{acknowledgments}
This work was supported by NSF grant CTS0625890, NSF-DMR 0907146, and NSF REU funding.  We thank Bruno Eckhard for valuable discussions.
 
\end{acknowledgments}

\end{document}